\documentclass[11pt]{article}

\usepackage{arxiv}

\usepackage[utf8]{inputenc} % allow utf-8 input
\usepackage[T1]{fontenc}    % use 8-bit T1 fonts

\usepackage{amsmath}
\usepackage{amssymb}
\usepackage{amsfonts}
\usepackage{mathtools}

\usepackage{booktabs}       % professional-quality tables
\usepackage{graphicx}
\usepackage{caption}
\usepackage{subcaption}
\usepackage{placeins}
\usepackage{setspace}
\usepackage{wrapfig} % for text-wrapping figures and tables

\usepackage{nicefrac}       % compact symbols for 1/2, etc.
\usepackage{microtype}      % microtypography
\usepackage{soul}           % highlighting
\usepackage{xcolor}         % colors
\usepackage{url}            % simple URL typesetting

\usepackage{natbib}
\usepackage{doi}

\usepackage{hyperref}       % hyperlinks
\usepackage{cleveref}       % smart cross-referencing (after amsmath + hyperref)

\usepackage{lipsum}         % remove once you have real content

\newcommand\BibTeX{{\rmfamily B\kern-.05em \textsc{i\kern-.025em b}\kern-.08em
T\kern-.1667em\lower.7ex\hbox{E}\kern-.125emX}}

\newcommand{\halfwidth}{0.48\textwidth}
\DeclareMathOperator*{\argmin}{arg\,min}

\title{Modeling diesel output particulate matter as the Ornstein-Uhlenbeck process}

\author{
  Maxwell Bolt \\
  School of Mechanical Engineering, Purdue University, IN, USA \\
  \AND
  Alex Alberts \\
  School of Mechanical Engineering, Purdue University, IN, USA \\
  \AND
  Akash S. Desai \\
  Cummins Techincal Center, Cummins Inc., Columbus, IN, USA \\
  \AND
  Peter Meckl \\
  School of Mechanical Engineering, Purdue University, IN, USA \\
  \AND
  Ilias Bilionis\thanks{Corresponding author: \texttt{ibilion@purdue.edu}} \\
  School of Mechanical Engineering, Purdue University, IN, USA
}

\hypersetup{
pdftitle={Modeling diesel output particulate matter as the Ornstein-Uhlenbeck process},
pdfsubject={},
pdfauthor={Maxwell Bolt, Alex Alberts, Akash Desai, Peter Meckl, Ilias Bilionis},
pdfkeywords={Particulate matter, Diesel compression ignition engines, Exhaust emissions, Ornstein-Uhlenbeck process},
}

\date{March 12, 2026}

\begin{document}
\maketitle

\begin{abstract}
Diesel engine particulate matter (PM) is one of the most challenging emission constituents to predict.
As engines become cleaner and emissions levels drop, manufacturers need reliable methods to quantify the PM generated by production engines. 
Due to the inaccuracy of commercial-grade sensors, they turn to predictive models to accurately estimate PM. 
In practice, this requires a computationally inexpensive model that provides PM estimates with calibrated uncertainty.
Complex, multiscale physics make mechanistic models intractable and traditional data-driven methods struggle in transient drive cycles due to the stochastic nature of PM generation.
Leveraging recent innovations in PM measurement technology, we introduce a novel PM model based on the Ornstein-Uhlenbeck (OU) process.
The OU process is a mean-reverting stochastic process commonly used in financial modeling, now being explored for engineering applications, and can be described as a stochastic differential equation (SDE).
We modify the OU process by parameterizing the terms of the SDE as functions of the engine state, which are then fit with a maximum likelihood estimate. 
In a synthetic example, we verify the ability of our model to learn a time-varying, parametrized OU process.
We then train the model using real experimental data designed to dynamically cover the engine operating space and test the trained model on EPA-regulated drive cycles. 
For most drive cycles, we find the method accurately predicts cumulative output of PM across time.
\end{abstract}

\onehalfspacing

% keywords can be removed
\keywords{Particulate matter \and Diesel compression ignition engines \and Exhaust emissions \and Ornstein-Uhlenbeck process}

\section{Introduction}
\label{sec:introduction}
Diesel particulate matter (PM) output poses significant public health and environmental concerns. In response, engine manufacturers and regulators continue to reduce tailpipe emissions for on-highway vehicles~\cite{epa2024regs}, with European regulations soon requiring continuous emissions reporting throughout an engine's life~\cite{eu2024reg1257}. 
While commercially available PM sensors are often adequate for diagnostics, high-fidelity predictive models allow manufacturers to closely monitor product health.
As these engines continue to become less polluting, a new generation of models is needed to characterize the stochastic behavior of the low PM regime.

An integral part of predicting tailpipe emissions is understanding engine-out emissions. 
While much of the tailpipe PM is mitigated by the after-treatment system, models of tailpipe PM rely on engine-out PM as the key input. 
Additionally, models of engine-out PM can be used to predict and mitigate after-treatment warranty issues, especially diesel particulate filter clogging~\cite{Zollner2021}.
Change in an engine's long-horizon PM output can indicate degradation in the fuel system, making engine-out PM a valuable prognostic tool. 
Throughout the rest of this work, we only concern ourselves with engine-out PM modeling, and all further discussion of PM refers to engine-out PM. 

PM generation is a complex chemical process primarily determined by in-cylinder states. 
Formation and oxidation occur on small time scales at the flame diffusion front~\cite{heywood1988} before propagating downstream to the measurement location at the turbine inlet, where we call the measured quantity PM generation. 
As PM generation is spatially dependent, manufacturing tolerances in parts such as the fuel injectors can lead to observable variation in output PM from engine to engine and cylinder to cylinder~\cite{LEACH2018340}. 
Furthermore, exhaust gas recirculation (EGR) changes the composition and thermodynamic state of the input charge gas, resulting in combustion event variation in the measured PM, even at steady-state operating conditions~\cite{Lee2008}. 
The physical and chemical processes generating PM occur in a highly dynamic diffusion flame that we cannot directly observe, and the lab-grade PM sensor itself has limited fidelity; as a result, the engine-out PM signal exhibits strong variability and should be modeled as a stochastic process.

Because of these complex phenomena present in internal combustion modeling, semi-empirical models have historically been used for PM modeling~\cite{hiroyasu1976}.
With the rise of machine learning, data-driven methods have become very popular for modeling internal combustion engines~\cite{ALIRAMEZANI2022100967}. Shahpouri et al.~\cite{shahpouri2021hybrid} compare model architectures such as regression trees, ensemble regression trees, support vector machines, Gaussian process regression (GPR), neural networks, and Bayesian neural networks. 
The architectures are compared using various selections of experimental and simulation data at set points on the torque map. The comparison concluded that GPR was the optimal architecture for modeling PM. 

In order to obtain a more robust GPR model, we implemented a Gaussian process (GP) with a structured deep kernel~\cite{wilson2016deep} to capture time dependencies, similar to the NOx model in Zinage et al.~\cite{zinage2025causal}.
However, we observed poor performance on test datasets, which led us to inspect the model's standardized errors.
We identified that the errors were correlated with each other, introducing model error traces. 
In regression problems, we expect model errors to be uncorrelated and appear as a random scatter~\cite{Draper1998}.
The conclusion that follows is that in transient modeling, regression is an incorrect approach; a better way is to model a drive cycle's PM dynamics with a physics-based model.

Very little transient modeling of PM exists in the literature, despite the Federal Test Procedure (FTP) cycle being a transient, federally regulated drive cycle.
Brahma et al.~\cite{Brahma2009} provided an excellent foundation in transient emissions modeling, where they performed regression using a generalized linear model. 
We aim to improve upon their work, noting that since its publication,  lab-grade sensors can directly measure PM output rather than estimate it from smoke opacity.

After identifying shortcomings of regression and the gap in transient modeling PM literature, we explored various established data-driven methods for modeling dynamical systems. 
Sparse identification of nonlinear dynamical systems (SINDy)~\cite{Brunton2016} is one popular method in the fluids community due to its flexible basis functions and model compactness, a valuable trait for production-grade engine control modules with limited memory.
Despite the advantages that SINDy provides, it has a significant limitation in that it needs derivatives of the input and output signals.
To train a SINDy-based PM model from experimental data, we  numerically approximated the derivatives through a variety of methods, all of which introduced numerical errors and led to poor test results.
Another popular data-drive method is neural ordinary differential equations (neural ODEs)~\cite{NODE}, a method where the dynamics are parametrized with a neural network. But because of the increased flexibility and number of parameters in neural networks, the model would overfit and quickly diverge when predicting on test datasets. 

In experimental datasets, we observe cycle-to-cycle variation by measuring different amounts of PM at the same operating condition, leading us to the conclusion that PM generation is best modeled as a stochastic process.
To effectively capture the stochastic dynamics of PM, we describe PM generation as a stochastic differential equation (SDE)~\cite{Oksendal2003}, a form which expresses deterministic dynamics as well as random fluctuations. 
By modeling dynamics, we can capture how PM generation results from changes in the engine state, and by modeling the random fluctuation, we account for the observed cycle-to-cycle variation.
Separating the model components into distinct drift and volatility terms show how different engine states drive the different characteristics of PM generation.

The contribution of this paper includes introducing SDEs as a framework for modeling complex phenomena in diesel engines. 
We specifically leverage the Ornstein-Uhlenbeck (OU) process for its mean-reverting nature, a process historically used in financial modeling, known as the Vasicek model~\cite{VASICEK1977177}, and more recently been used in physical modeling~\cite{FRONDELIUS2022104454}. 
We theorize that the engine operating conditions set an average PM level, then through gas transport, EGR, and sensor delays, the output PM approaches the level set by the mean. 

Palamarchuk investigates the properties of the OU process with time-dependent coefficients~\cite{Palamarchuk2018}. 
Our work extends time-dependent coefficients by parameterizing the drift and volatility as functions of the engine state in order to model PM as a continuously shifting OU process.
The method allows us to experiment with different functions parameterizing the stochastic process terms, functions whose parameters we fit with a maximum likelihood estimate (MLE). 
Once trained, we can sample from this SDE and characterize PM generation from the sample path statistics. 
With this method, we generate a probability distribution of PM mass flow at each time step, informed by the previous prediction and the engine state.

This paper will first outline in the Methodology section our methods, training, and performance metrics. 
Then, we demonstrate the methodology on a synthetic example in the Model Verification section.
Next, we compare our model's performance to experimental EPA-defined drive cycles and benchmark it against various model architectures. 
Finally, we discuss the implications of our findings and suggestions for further research.

\section{Methodology}
\label{sec:methodology}

\subsection{Model Development}
\label{sec:model-development}
At steady-state operating points, modern engines produce very small amounts of PM. 
In transient duty cycles, however, PM emission spikes are often an order of magnitude larger than steady-state emissions. 
To accurately predict PM emissions and ensure regulatory compliance, engine manufacturers must account for these transient regimes. 

Because of the complex chemical processes of in-cylinder combustion and limited knowledge about the progression of the flame front, PM exhibits inherently stochastic behavior.
As a result, creating purely regression-based models for PM has proved difficult.
The difficulty of building a robust regression model is then compounded by the highly variable commanded engine signals, such as fuel injection and valve timing, and partially observed engine states, pressures, and temperatures at various points.

\begin{figure}[!htbp]          % try Here/Top/Bottom/Page
  \centering
  \includegraphics[width=0.7\textwidth]{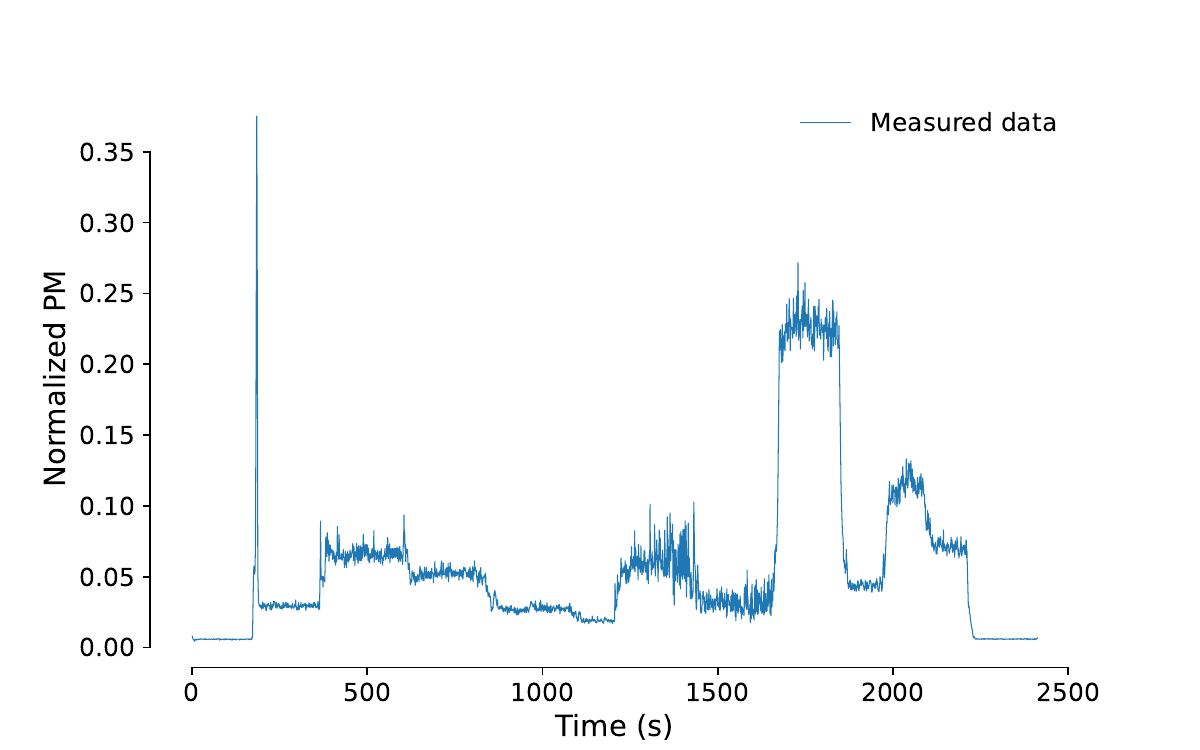}
  \caption{Measured PM generation during RMCSET cycle.}
  \label{fig:RMCSET_data}
\end{figure}

We formulated our PM dynamics model after observing the measured PM data of another EPA-defined drive cycle, the Ramped Mode Cycle Supplemental Emissions Test (RMCSET) as seen in Fig.~\ref{fig:RMCSET_data}. 
Whereas the FTP continuously moves through transient regimes, the RMCSET holds a steady-state operating condition, makes a transient move, and then holds a new steady-state operating condition.
From the RMCSET, we see that a set engine operating condition creates some average amount of PM with stochastic fluctuations around this mean value. 
When the engine moves to the next steady-state operating condition, there is a new PM mean corresponding to this set point, but the engine gasses and control state take some time to transition to this mean from the previous set point. 
We model this transition using the OU process, whose mean-reversion parameter characterizes the rate at which the PM signal approaches the new mean.

\subsection{Model Formulation}
\label{sec:model-formulation}
We model PM generation with a modification of the OU process, which can be described as an SDE in the It\^o sense~\cite{Oksendal2003}.
That is, letting $X_t$ represent the value of PM at time $t$, the OU process is governed according to the SDE
\begin{equation}
    \label{eqn:OU}
    dX_t = \lambda(\mu - X_t)dt + \sigma dW_t,
\end{equation}
with some initial state $X_0$.
The first term $\lambda(\mu - X_t)$ is known as the drift and models the deterministic component of the dynamics.
The parameters $\mu$ and $\lambda$ are, respectively, the mean value around which $X_t$ fluctuates and the mean reversion parameter, governing how quickly $X_t$  approaches the level set by the mean.
The $\sigma$ term is the volatility, which captures the stochasticity of the process, modulating the magnitude of the random fluctuations.
Here, $W_t$ represents a standard Brownian motion.

In transient operation, the engine operating condition will change dynamically.
Therefore, we model PM by modifying eq.~(\ref{eqn:OU}) to be
\begin{equation}
    \label{eqn:PMsde}
    dX_t = \lambda(\mu_t - X_t)dt + \sigma_t dW_t.
\end{equation}
This form captures the behavior that PM stochastically fluctuates around some level set by the engine state, which changes dynamically, given by $\mu_t$.
From the PM data, we expect that $\mu_t$ will be a simple piecewise constant function of time.
We also treat the noise standard deviation $\sigma_t$ as being time-dependent.

\clearpage % or \newpage (forces the next page)
\begin{wraptable}{r}{0.55\textwidth}
  \vspace{-1.2\baselineskip} % tweak vertical shift if needed
  \centering
  \caption{Engine variables used as model inputs.}
  \label{tab:engine-variables}
  \begin{tabular}{@{}cl@{}}
    \toprule
    \textbf{\#} & \textbf{Engine Variable} \\
    \midrule
    1  & Engine speed \\
    2  & Engine brake torque \\
    3  & Main injection timing \\
    4  & Main injection quantity \\
    5  & Pilot 2 injection timing \\
    6  & Pilot 2 injection quantity \\
    7  & Post 1 injection timing \\
    8  & Fuel rail pressure \\
    9  & Variable-geometry turbocharger valve actuation \\
    10 & EGR valve actuation \\
    11 & EGR outlet temperature \\
    12 & Intercooler outlet temperature \\
    13 & Mass flow of charge air \\
    14 & Mass flow of EGR \\
    15 & In-cylinder oxygen estimate \\
    16 & Turbine inlet temperature \\
    \bottomrule
  \end{tabular}
\end{wraptable}

To connect the current engine state to the PM, we treat the mean level $\mu_t$ as a parametrized function of the measured states and engine control inputs.
We chose the variables based on physical intuition, domain knowledge, and model experimentation.
The engine variables that we consider are found in Table~\ref{tab:engine-variables}.
We also treat the volatility $\sigma_t$ in the same way, describing the way that the engine states cause random PM fluctuations.
The mean reversion parameter $\lambda$ is treated as a constant, and it quantifies the time delay to reach the level set by the current operating point.
Our model also assumes that PM is first-order Markovian, meaning that the PM dynamics are determined only by the previous time step.

In all our datasets $\mu_t$ and $\sigma_t$ appear as piecewise constant functions of time.
The physics of PM work such that the map between the engine state and these signals results in a sum of simple functions.
We choose to linearly parameterize the mean function, $\mu_t$, and the volatility scaling, $\sigma_t$, effectively taking the Taylor series of the engine state to signal map.
In numerical experiments not reported in this paper, we had parametrized this map as a deep neural network, but struggled with overfitting to the training data.
So, in order to not overfit and have a compact model, we work with the linear parameterization.

Letting $U_i$, $i=1,\dots,16$ denote the individual engine variables, we model the mean level of PM according to
\begin{equation*}
    \mu_t = \sum_{i=1}^{16}a_iU_{i,t} + b,
\end{equation*}
where $(a_i)_{i=1}^{16}$ and $b$ are trainable parameters.
To enforce the fact that the volatility must be positive, we parametrize it linearly before passing it through a softplus transformation,
\begin{equation*}
    \sigma_t = \log \left(1 + \exp\left(\sum_{i=1}^{16}c_iU_{i,t} + d\right) \right),
\end{equation*}
where $(c_i)_{i=1}^{16}$ and $d$ are the model parameters for $\sigma_t$.
The mean reversion parameter $\lambda$ is a single learned constant for all times and all duty cycles.
We let $\theta = (a_1,\dots,a_{16},c_1,\dots,c_{16},b,d,\lambda)$ denote the collection of all trainable parameters.
Plugging in our parameterization into eq.~(\ref{eqn:PMsde}), we arrive at the PM model used in this work,
\begin{equation}
    \label{eqn:SDEmodel}
    dX_t(\theta) = \lambda(\mu_t(\theta) - X_t(\theta))dt + \sigma_t(\theta)dW_t.
\end{equation}

\subsection{Model Training}
\label{sec:model-training}
We use the maximum likelihood approach to optimize the parameters $\theta$.
This requires characterizing a sample path $X_t(\theta)$ to evaluate the model likelihood during each iteration. In other words, we must numerically solve the SDE in some time window.
We initially worked with the Euler-Maruyama scheme~\cite{kloeden1977numerical}. 
However, we had limited success using the Euler-Maruyama scheme, possibly from the size of our data sampling time step causing instabilities.

Instead, we capture the solution to the SDE using the Fokker-Planck representation.
The Fokker-Planck representation describes the time-evolution of the solution to eq.~(\ref{eqn:SDEmodel}) as a probability density function (PDF) $p_{\theta}(x,t)$.
This PDF is governed by the Fokker-Planck equation
\begin{equation}    
    \label{eqn:FP}
    \frac{\partial p_{\theta}}{\partial t} = -\frac{\partial}{\partial x}\left\{\lambda(\mu_t(\theta) - x)p_{\theta}(x,t)\right\} + \frac{1}{2}\sigma_t^2(\theta)\frac{\partial^2 p_{\theta}}{\partial x^2}.
\end{equation}
The Green's function of eq.~(\ref{eqn:FP}) encodes the transition probability of a state $X_t(\theta)$ moving to $X_{t+\Delta t}(\theta)$ for a given time step $\Delta t$.
For the OU process, it is well-known that the transition probability will be Gaussian.

We can identify the mean and variance of the transition probability by taking expectations of eq.~(\ref{eqn:PMsde}).
Without loss of generality, we study the transition to a state $X_t(\theta)$ from the initial condition $X_0$ to simplify the mathematics that follow.
For the mean, which we denote by $m_{\theta}$, we find
\begin{align*}
    \mathbb{E}[dX_t(\theta) \mid X_0]
    &= \mathbb{E}\big[\lambda \big(\mu_t(\theta)-X_t(\theta)\big) dt \mid X_0\big]
       + \mathbb{E}\big[\sigma_t(\theta)\,dW_t \mid X_0\big] \\
    &= \lambda\big(\mu_t(\theta)-m_{\theta}\big) dt.
\end{align*}
where the noise term vanishes since it is zero-mean Gaussian.
Moving the $dt$ over to the left-hand side and evaluating the final expectation, we find an ODE for the mean,
\begin{equation}
    \label{eqn:meanode}
    \frac{dm_{\theta}}{dt} = \lambda(\mu(t;\theta) - m_{\theta}), \quad m_{\theta}(0) = X_0.
\end{equation}
Eq.~(\ref{eqn:meanode}) is easy to solve by integration:
\begin{equation}
    \label{eqn:mean}
    m_{\theta}(t) = X_0e^{-\lambda t} + \lambda \int_0^t \mu_s({\theta})e^{-\lambda(t-s)}ds.
\end{equation}

The transition variance follows a similar pattern, where we find we must evaluate
\begin{equation}
    \label{eqn:var}
    V_{\theta}(t) = \int_{0}^t\sigma_s^2(\theta)e^{-2\lambda(t-s)}ds.
\end{equation}
Before proceeding further, we have two comments to make on eqs.~(\ref{eqn:mean}) and~(\ref{eqn:var}).
First, observe that the kernel appearing in both integrals $e^{-\lambda(t-s)}$ is exactly the covariance kernel of the OU process.
Secondly, if $\mu_t(\theta)$ and $\sigma_t(\theta)$ are constants, the integrals can be evaluated analytically as
\begin{equation}
    \label{eqn:mean_var}
    \left\{
    \begin{split}
    m_{\theta}(t) &= X_0e^{-\lambda t} + \mu_t(\theta)\left(1 - e^{-\lambda t}\right), \\
    V_{\theta}(t) &= \frac{\sigma_t^2(\theta)}{2\lambda}\left(1 - e^{-2\lambda t}\right),
    \end{split}
    \right.
\end{equation}
which revert to exactly the transition mean and variance of the OU process.

Now, recall that in our model $\mu_t(\theta)$ is assumed to be a sum of simple functions, i.e., the mean level of PM set by the engine commands.
Let $t_0=0,t_1,\dots,t_{f-1},t_f=T$ denote the location of the jumps in $\mu_t(\theta)$.
Using eq.~(\ref{eqn:mean}), we may write for the mean at an arbitrary time step, conditional on the previous one, as
$$
m_{\theta}\left(t_{i+1}\right) = X_{t_i}e^{-\lambda \Delta t} + \mu_{t_i}(\theta)\left(1-e^{-\lambda \Delta t}\right).
$$
The same is true for $V_{\theta}$, as $\sigma_t(\theta)$ is also a piecewise constant function.
Hence, we see that the transition mean and variance are indeed given by eq.~(\ref{eqn:mean_var}), where both $\mu_t(\theta)$ and $\sigma_t(\theta)$ are piecewise constant functions, set by the engine signals.

Therefore, for an arbitrary time $t$ and transition length $\Delta t$, we are justified in writing the transition probability as
\begin{equation}
  \label{eqn:trans}
  p\left(X_{t+\Delta t}(\theta)\middle|X_t(\theta)\right)
  = \frac{1}{\sqrt{2\pi\,V_{\theta}(t)}}
    \exp\left(
      -\frac{(X_{t+\Delta t}-m_{\theta}(t))^2}{2V_{\theta}(t)}
    \right).
\end{equation}
To define the likelihood for this problem, we use the above transition probability.
Let $(t_i)_{i=0}^N$ denote the time steps where the data are collected and $(Y_i)_{i=0}^N$ be the corresponding measurement of PM.
Then, for an individual measurement $(t_i,Y_i)$, the likelihood is given by inserting the pair into eq.~(\ref{eqn:trans}), returning the Gaussian:
$$
p(Y_i|Y_{i-1},\theta) = \mathcal{N}(Y_i|m_{\theta}(t_i), V_{\theta}(t_i)^2).
$$
The joint likelihood between all measurements can be found using the product rule of probability:
\begin{equation*}
    p(Y_1,\dots,Y_N | Y_0,\theta)= \prod_{i=1}^Np(Y_i|Y_{i-1},\theta).
\end{equation*}

To train the model, we take the MLE.
That is, we identify the best parameters $\theta^*$ by solving the following equivalent optimization problem found by minimizing the negative $\log$-likelihood
\begin{equation}
  \label{eqn:loss}
  \theta^* = \argmin_{\theta} \sum_{i=1}^N \left\{
    \log\big(V_{\theta}(t_i)\big)
    + \frac{\big(Y_i - m_{\theta}(t_i)\big)^2}{V_{\theta}(t_i)}
  \right\}.
\end{equation}
When training the model according to the loss function eq.~(\ref{eqn:loss}), we minibatch the data and optimize the parameters $\theta$ using Adam~\cite{kingma2017}.
Using our framework, any arbitrary function can be chosen to parametrize the SDE time-varying coefficients, the parameters of which can be found via eq.~(\ref{eqn:loss}).
Of course, one would need to update the transition probability by evaluating the integrals found in eqs.~(\ref{eqn:mean}) and~(\ref{eqn:var}).

\subsection{Data Transformation}
\label{sec:data-transformation}
Across all datasets the experimental PM data are extremely skewed, where occasional spikes where an order of magnitude larger than the majority of the data. 
To better train the PM model, we first divide the training data by its standard deviation, and then apply a piecewise $\log$-linear transformation on the scaled data, according to
$$
Y_t =
\begin{cases}
\log\bigl(Z_t\bigr), & Z_t < 1.0,\\
Z_t - 1.0,           & Z_t \ge 1.0.
\end{cases}
$$
where $Z_t$ is the rescaled PM.
The resulting data $Y_t$ is used to train the model.
This piecewise $\log$-linear transformation allows us to model the PM in the $\log$-space while allowing the large spikes to remain expressive.
We experimented with several transformations of PM including z-score normalization and a simple $\log$ scale. However, we found the $\log$-linear transformation to be the most robust.

\subsection{Model Deployment}
\label{sec:model-deployment}
After training the model, we predict the time series of PM on new experimental datasets from the initial state at time $t=0$ to some nominal time $t=T$.
For a new dataset, let $(t_i)_{i=0}^N$ denote the time steps where each engine signal $U_i$ is measured.
We do not assume any value of PM is observed after training the model and predicting on new data.
Rather, we feed the trained PM model the mean function from the first time step as the initial condition.
That is, we take 
\begin{equation}
\label{eqn:init}
X_0 = \mu_{t_0} = \sum_{i=1}^{16}a_iU_{i,0} + b, 
\end{equation}
giving a deterministic initial PM value.
We find that the model is robust to choices of initial condition as it corrects itself using the measured engine states $U_i$ almost instantaneously.

Using the transition probability, we generate a prediction of PM at time $t_1$ by sampling eq.~(\ref{eqn:trans}).
A sample path is generated by iterating through this process until time $T$.
That is, we generate a sample path according to
\begin{equation}
    \label{eqn:predict}
    X_{t_{i+1}} = m_{\theta}(t_i) + \sqrt{V_{\theta}(t_i)} \eta_{t_i}, \quad i=0,\dots,N-1,
\end{equation}
where $\eta_{t_i} \overset{\mathrm{i.i.d.}}{\sim} N(0, 1)$.
We repeat this process $M =10,000$ times to generate sufficient sample path statistics.

\subsection{Performance Metrics}
\label{sec:performance-metrics}
Because PM is a stochastic process, it is difficult to determine when a model is a good fit by simply comparing an individual sample path to the data.
To evaluate the effectiveness of our model, we do not compare the predictions to the exact measured PM, but rather we assess whether or not the PM data reasonably falls within the probability distribution generated by the model. 
We can determine if the data are generated from the model distribution with the probability integral transform (PIT), also known as the inverse cumulative distribution function (CDF) test. 
The PIT compares the standardized errors between the model predictions and the data to the statistics of the standard normal distribution.
Let $X_{t_i}^1,\dots,X_{t_i}^M$ denote the samples generated by the model at an arbitrary time $t_i$.
We then evaluate the standardized error $e_{t_i}$ according to
$$
e_{t_i} = \frac{Y_{t_i}-\bar{X}_{t_i}}{s_i},
$$
where $\bar{X}_{t_i}$ and $s_i$ denote the empirical mean and standard deviation of the model at time $t_i$, respectively.
We then pass each $e_{t_i}$ through the CDF of the standard normal distribution, which we refer to as the PIT value.
If our data belong to the distribution generated by our model, the resulting PIT values would follow the uniform distribution. 

We use the Kolmogorov-Smirnov (KS) statistic to quantitatively measure how closely the PIT values are to being uniformly distributed. 
The KS statistic finds the difference between two distributions by calculating the supremum of the distance between two CDFs and is bounded between $0$ and $1$. 
If the PIT values are perfectly uniformly distributed, the KS statistic would be zero.

We can also qualitatively assess the model fit with quantile-quantile (Q-Q) plots, where we compare the empirical quantiles from the PIT values with the theoretical quantiles of the uniform distribution.
Two equivalent distributions create a Q-Q plot that lies perfectly on the 45-degree line.
Deviations from this line encode valuable information about the discrepancy between the predictions and the experimental data.

\begin{figure}[htbp]
  \centering
  \begin{subfigure}{\halfwidth}
    \centering
    \includegraphics[width=\linewidth]{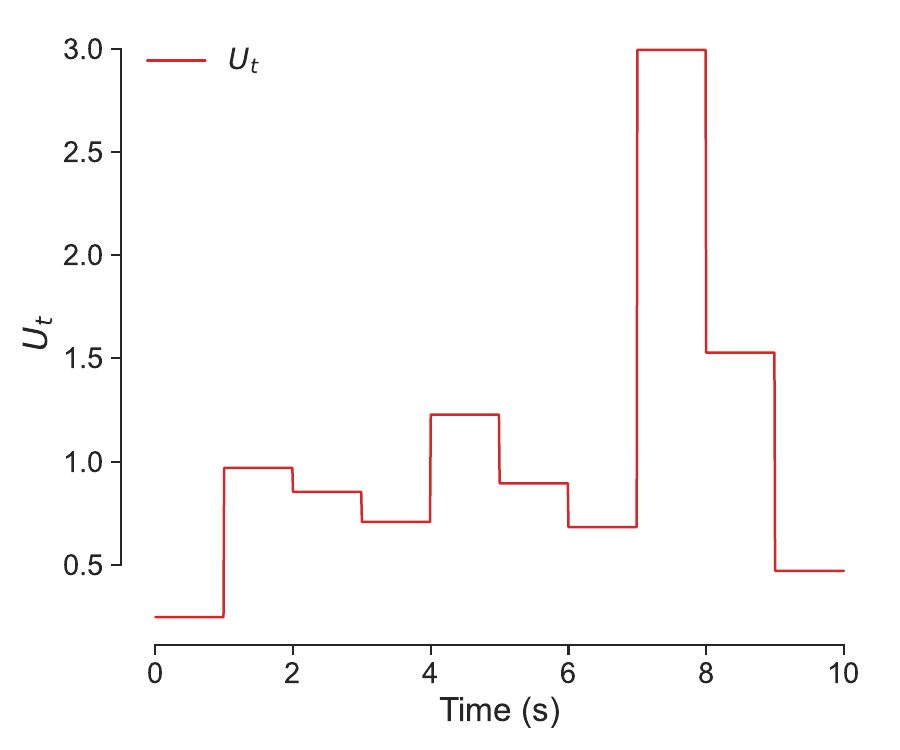}
    \caption{Synthetic example piecewise constant control}
    \label{fig:synth_U}
  \end{subfigure}
  \hfill
  \begin{subfigure}{\halfwidth}
    \centering
    \includegraphics[width=\linewidth]{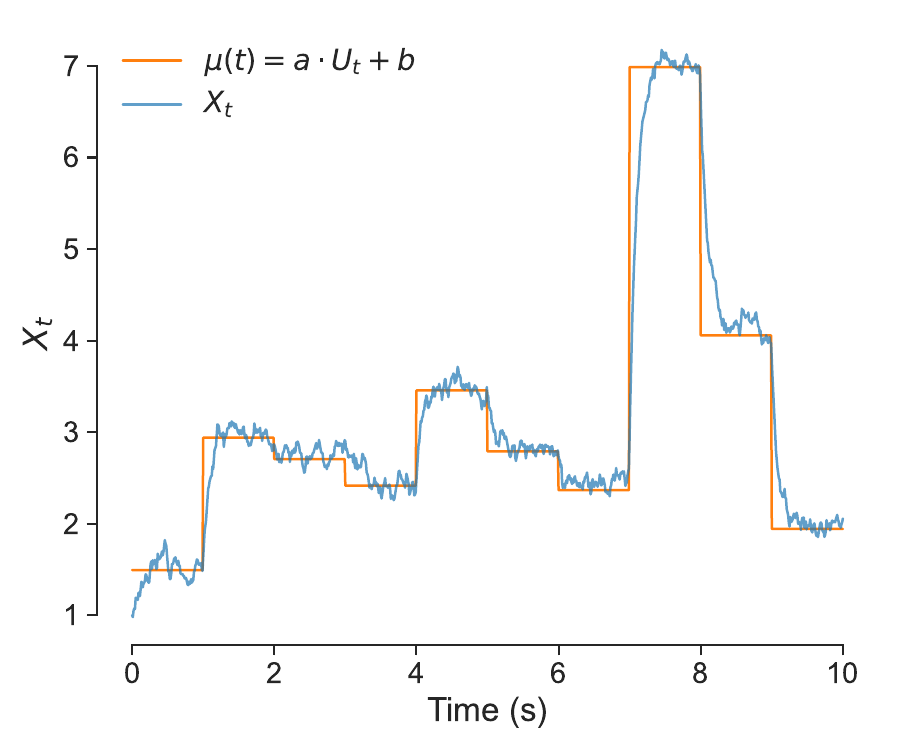}
    \caption{Synthetic example mean function and data}
    \label{fig:synth_mu}
  \end{subfigure}

  \caption{Synthetic example: (a) piecewise constant control input over time; 
  (b) resulting latent mean response and observations over the same time horizon.}
  \label{fig:synth_combined}
\end{figure}

\section{Model Verification}
\label{sec:model-verification}
To verify the method, we study a synthetic example generated with a single piecewise constant control $U_t$ as shown in Fig.~\ref{fig:synth_U}.
The latent state of the system responds linearly to the control.
This is similar to how we expect the engine controls and input to affect the mass of PM generated.
The mean function is then characterized as a linear function of $U_t$
\begin{equation*}
    \mu_t = aU_t+b,
\end{equation*}
where $a$ and $b$ are learnable
parameters.
The mean reversion, $\lambda$, and the volatility, $\sigma$, are treated as constants.
The synthetic data used to fit the parameters is shown in Fig.~\ref{fig:synth_mu}.
We train the model by observing the first 700 data points or $7$ seconds (just before the PM spike) and fitting the parameters with the MLE.

\begin{table}[htbp]
  \centering
  \caption{Synthetic example parameter estimates}
  \label{tab:param-estimates}
  \begin{tabular}{lccc}
    \toprule
    \textbf{Parameter} & \textbf{True} & \textbf{Estimated} & \textbf{\% Error} \\
    \midrule
    $\lambda$ & 10.0 & 10.963 & 9.63 \\
    $a$        & 2.0 & 1.994 & 0.32 \\
    $b$        & 1.0 & 1.065 & 6.48 \\
    $\sigma$   & 0.5 & 0.523 & 4.60\\
    \bottomrule
  \end{tabular}
\end{table}

The results are summarized in Table~\ref{tab:param-estimates}.
It appears that the most difficult parameter to fit is the mean reversion parameter $\lambda$.
If more data are observed or in cases with lower volatility $\sigma$, we find that the identification of $\lambda$ improves.
Despite a larger error in this parameter, our model identifies the remaining parameters with little error.

\begin{figure}[!htbp]          
  \centering
  \includegraphics[width=0.7\textwidth]{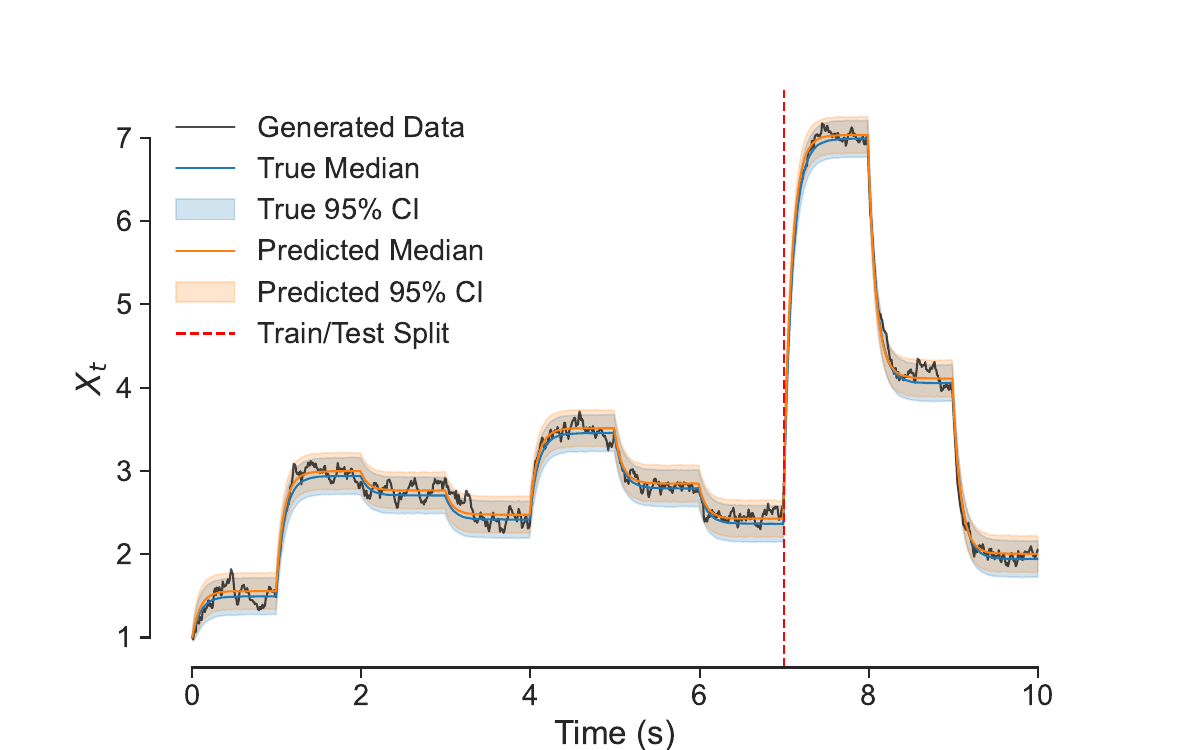}
  \caption{Synthetic example time series prediction}
  \label{fig:pred_dist}
\end{figure}
\clearpage
\begin{wrapfigure}{r}{0.4\textwidth}
  \vspace{-1.0\baselineskip} % tweak or remove if the vertical alignment looks off
  \centering
  \includegraphics[width=0.4\textwidth]{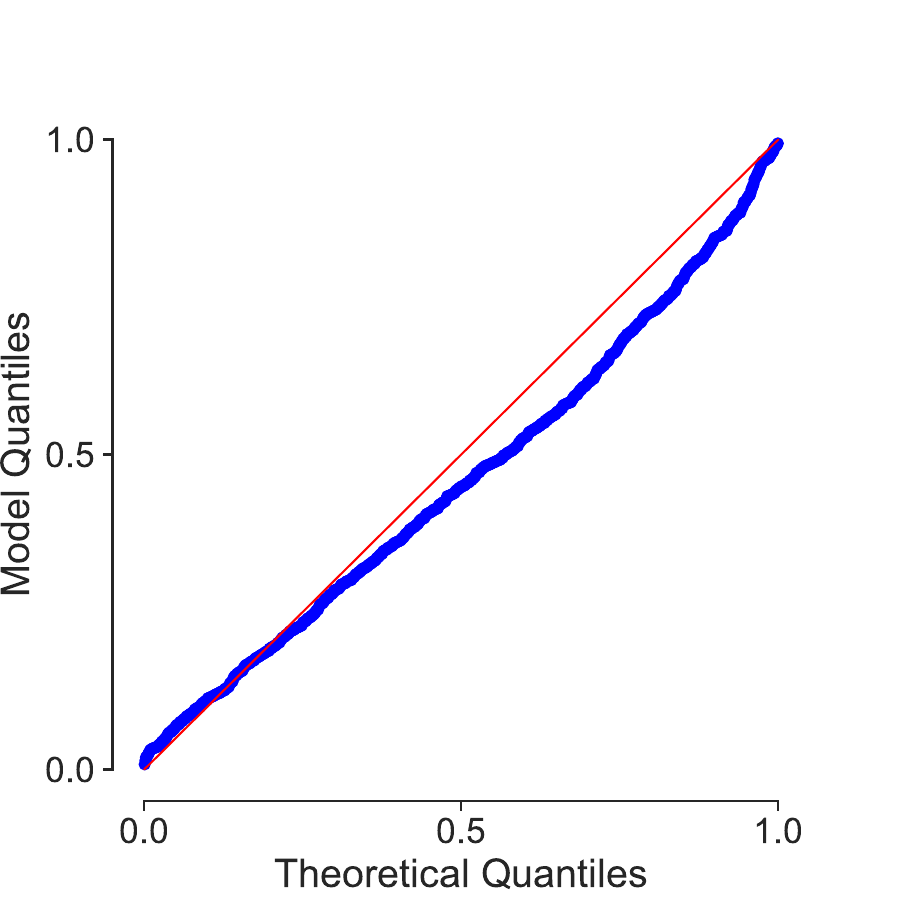}
  \caption{Synthetic example Q--Q plot.}
  \label{fig:synth_qq}
\end{wrapfigure}
After fitting the parameters with the MLE, we generate $10,000$ sample paths and collect the model statistics, which can be found in Fig.~\ref{fig:pred_dist}.
We display the median of the samples and use the quantiles to generate the $95\%$ credible interval (CI) to clearly show the distribution, and we find the model accurately predicts the entire dataset, including the last $3$ seconds after the PM spike where it was not trained.
The KS statistic from the inverse CDF test is 0.091, quantitatively showing that the data fits within the distribution generated by the model.
We also report the Q-Q plot of the PIT values in Fig.~\ref{fig:synth_qq} where we see generally close agreement with the uniform distribution reference quantiles.
At the upper quantiles, the Q-Q plot falls slightly below the 45-degree line, indicating that the observations tend to lie in the left tail of the predictive distribution, so our model is mildly overpredicting.
The code for this synthetic example can be found in the GitHub repository \url{https://github.com/PredictiveScienceLab/ou-pm-paper}.

\section{Results}
\label{sec:results}
\subsection{Experimental data and training procedure}
\label{sec:results-data-training}
The experimental data in this paper are from a Cummins mid-range B6.7L inline 6-cylinder diesel engine tested at the Cummins Technical Center in Columbus, IN. 
The reported PM quantities are mass flow rates measured by lab-grade sensors and normalized by the same global maximum across all datasets.
The engine variables used as inputs are measured using a mixture of both lab-grade and production-grade sensors. 
The model was trained on several in-house drive cycles designed to explore all possible operating conditions by systematically varying each engine control, then tested on Cummins and EPA-designed duty cycles.

We identified a subset of our available training data which covers a large area of the torque map, withheld it from training, and used it to do early stopping~\cite{Goodfellow-et-al-2016}.
We will refer to this subset of our data as our validation dataset which equates to $15\%$ of our total available training data.
The model is implemented using the JAX Python library~\cite{jax2018github} in combination with Equinox~\cite{kidger2021equinox}. 
Optimization is performed using the Adam algorithm~\cite{kingma2017} as implemented in Optax~\cite{deepmind2020jax}.
We train the model for 1000 epochs, which takes roughly $135$ seconds on a MacBook Pro equipped with the Apple M2 Max chip and 64 GB of memory.
We then keep the parameters that performs best on the validation data and save them as our final model.

\subsection{Test-cycle performance}
\label{sec:results-global}
Cummins provided us with six test datasets that they use to assess model performance.
Four of them are variants of the FTP and RMCSET cycles that the EPA uses to regulate emissions output.
The remaining two are Cummins designed snap throttle cycles, denoted STC.
For each test dataset, we feed the trained model an initial PM value from eq.~(\ref{eqn:init}), and use eq.~(\ref{eqn:predict}) to generate sample paths of PM.
The KS statistic resulting from the inverse CDF test for each test dataset is listed in Table \ref{tab:ks-statistics}.

\begin{table}[!htbp]
  \centering
  \caption{Kolmogorov–Smirnov statistics for test datasets.}
  \label{tab:ks-statistics}
  \begin{tabular}{lc}
    \toprule
    \textbf{Test Dataset} & \textbf{KS Statistic} \\
    \midrule
    FTP - Low NOx           & 0.085 \\
    FTP - Nominal NOx       & 0.290 \\
    FTP - High NOx          & 0.246 \\
    RMCSET - Nominal NOx    & 0.482 \\
    STC - Nominal NOx    & 0.134 \\
    STC - High NOx       & 0.057 \\
    \bottomrule
  \end{tabular}
\end{table}

\begin{figure*}[!htbp]
    \centering
    % Row 1 (four across)
    \begin{subfigure}[b]{0.2375\textwidth}
        \centering
        \includegraphics[width=\textwidth]{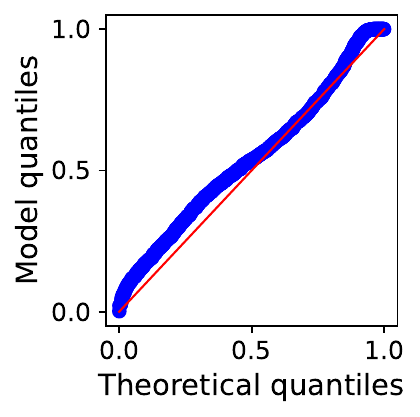}
        \caption{FTP - Low NOx.}
        \label{fig:qq_ftp}
    \end{subfigure}
    \begin{subfigure}[b]{0.2375\textwidth}
        \centering
        \includegraphics[width=\textwidth]{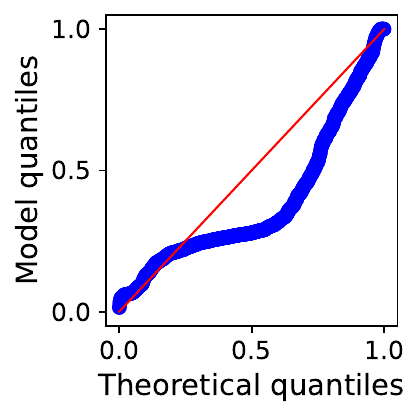}
        \caption{FTP - Nominal NOx.}
    \end{subfigure}
    \begin{subfigure}[b]{0.2375\textwidth}
        \centering
        \includegraphics[width=\textwidth]{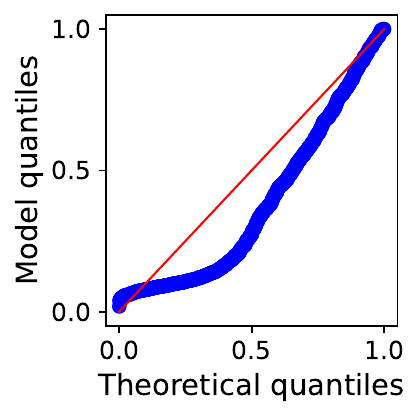}
        \caption{FTP - High NOx.}
    \end{subfigure}
    \begin{subfigure}[b]{0.2375\textwidth}
        \centering
        \includegraphics[width=\textwidth]{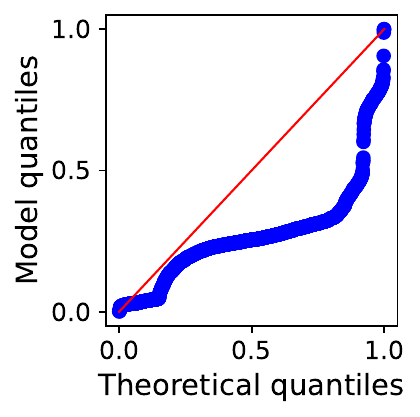}
        \caption{RMCSET - Nominal NOx.}
        \label{fig:qq_rmcset}
    \end{subfigure}

    % Row 2 (under the middle two of row 1)
    \begin{subfigure}[b]{0.2375\textwidth}
        \vspace{0pt}
    \end{subfigure}
    \begin{subfigure}[b]{0.2375\textwidth}
        \centering
        \includegraphics[width=\textwidth]{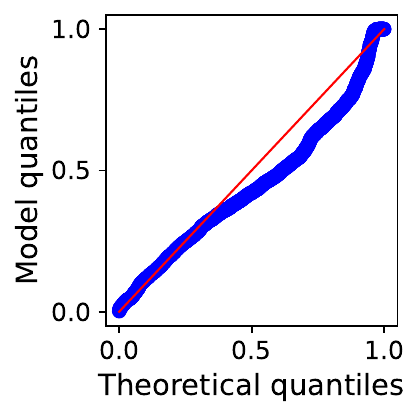}
        \caption{STC - Nominal NOx.}
    \end{subfigure}
    \begin{subfigure}[b]{0.2375\textwidth}
        \centering
        \includegraphics[width=\textwidth]{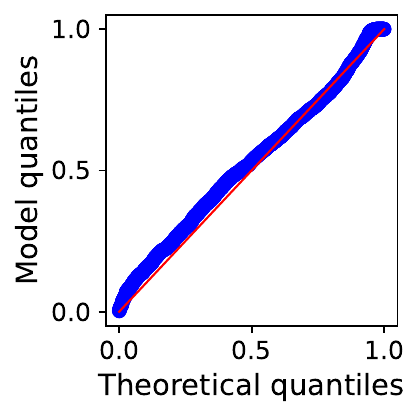}
        \caption{STC - High NOx}
    \end{subfigure}
    \begin{subfigure}[b]{0.2375\textwidth}
        \vspace{0pt}
    \end{subfigure}

    \caption{Q-Q plots for test datasets.}
    \label{fig:qq}
\end{figure*}

We find that in all cases except for one, namely the RMCSET cycle, the model performs well.
The Q-Q plots in Fig.~\ref{fig:qq} qualitatively compare the model's predictive distribution to the reference distribution for each test dataset.
In cases where the model needs improvement, the Q-Q plots suggest the same trend: that observations lie in the left-tail of the predictive distribution and the model consistently overpredicts PM.

\begin{figure}[!htbp]          
  \centering
  \includegraphics[width=0.5\textwidth]{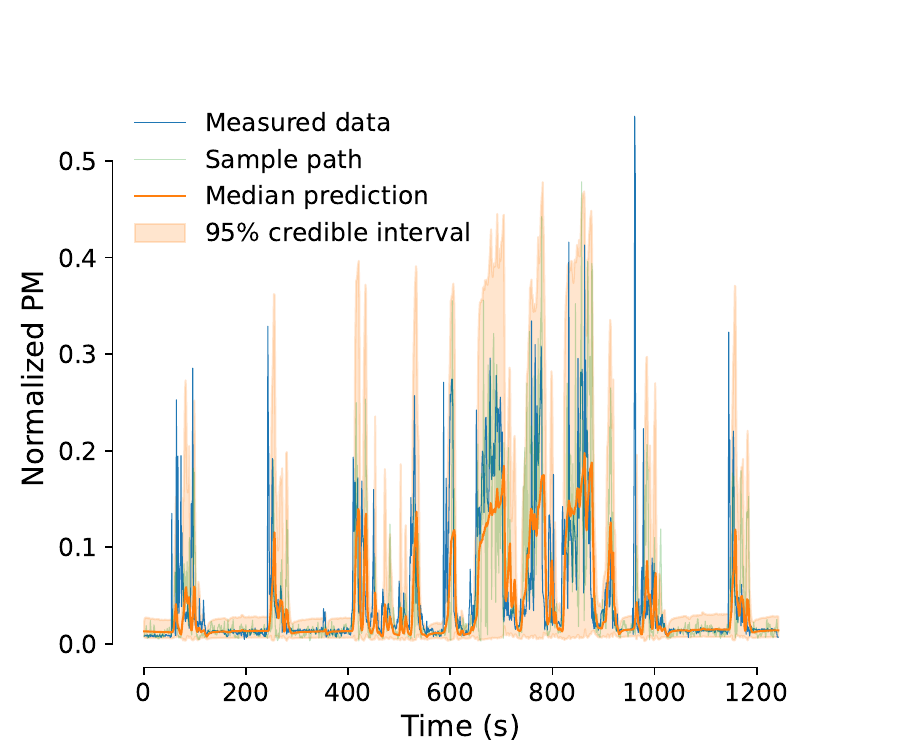}
  \caption{Time series prediction for FTP - Low NOx cycle}
  \label{fig:FTP_6}
\end{figure}

To demonstrate the performance of the model, we provide in-depth results for two test datasets and summarize the rest.
The first is the FTP - Low NOx dataset, which is the most commonly used regulator drive cycle. 
The time series plot of the FTP - Low NOx is given in Fig.~\ref{fig:FTP_6}, where we plot the measured data as well as the median prediction and the 95\% CI.
We also show a representative sample path generated by the model.
Qualitatively comparing the model sample path to the measured data shows the measured data can be thought of as a sample path from the model in agreement with the low KS statistic.
From the time series plot, we see that the model moves well from one PM level set to another but struggles to capture some of the more dramatic spikes.
% time series
From the KS statistic and Q-Q plots, we see the model performs the worst on the RMCSET cycle.
A probable explanation for the model's poor performance in this case is that the training data transiently varies the engine controls, whereas the RMCSET explores moving between set steady-state operating conditions.
Hence, this dataset is likely out of distribution.

\begin{figure}[htbp]
  \centering
  \begin{subfigure}{\halfwidth}
    \centering
    \includegraphics[width=\linewidth]{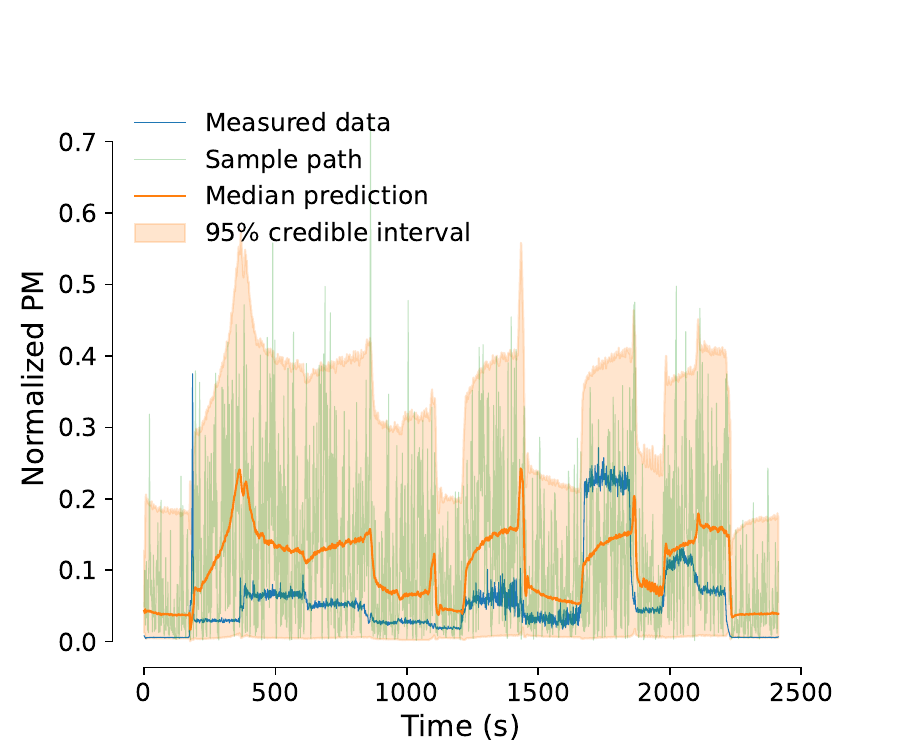}
    \caption{Time series prediction.}
    \label{fig:RMCSET_ts}
  \end{subfigure}
  \hfill
  \begin{subfigure}{\halfwidth}
    \centering
    \includegraphics[width=\linewidth]{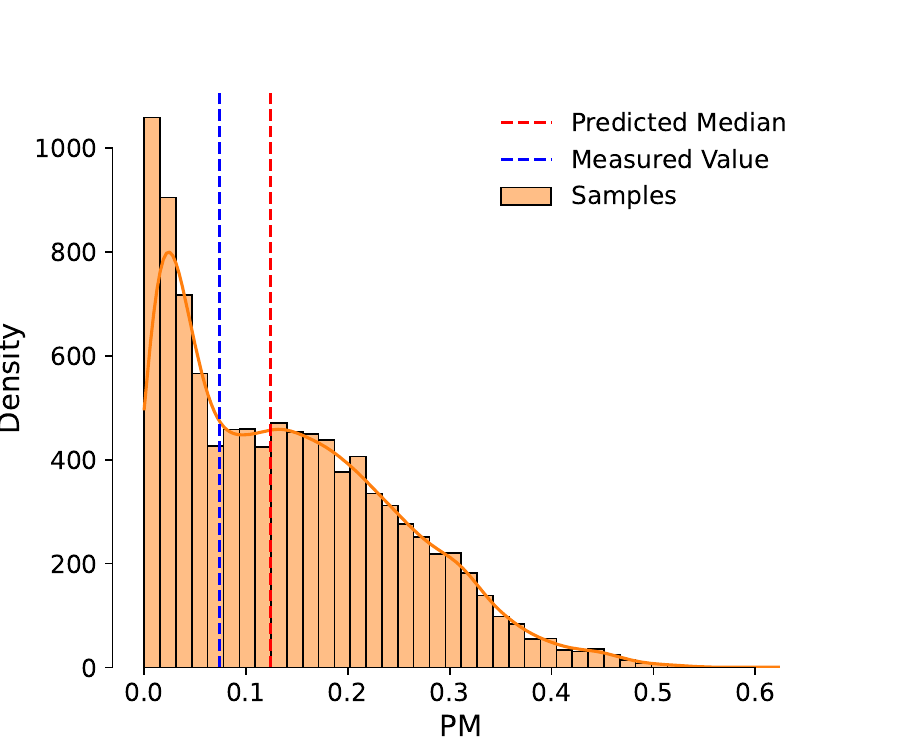}
    \caption{Predictive distribution at $t = 604\,\mathrm{s}$.}
    \label{fig:RMCSET_dist}
  \end{subfigure}
  \caption{Model performance on the RMCSET cycle: (a) time series prediction; (b) predictive distribution at a representative time.}
  \label{fig:RMCSET_perf}
\end{figure}
From the time series plot seen in Fig.~\ref{fig:RMCSET_ts}, we confirm this by noting that our individual sample path does not resemble the measured data.
Despite poor performance, the measured time series data stays within the CI.
Moreover, by studying the CI we can better understand the model's faults.
Fig.~\ref{fig:RMCSET_dist} shows the predictive distribution at a randomly chosen instant in time.
We observe that the distribution generated by the model is heavily skewed towards larger values of PM, which causes the median to overpredict. 
Also of note is that there is significant probability mass near the measured data, meaning that it is still probable that this PM value can reasonably be sampled from the model.
Better performance on the RMCSET could be achieved by including steady-state transition cycles within the training data or by including a prior/regularization term in the model to increase the sharpness of the resulting distribution.

\begin{figure*}[!htbp]
    \centering
    \begin{subfigure}[b]{0.3175\textwidth}
        \centering
        \includegraphics[width=\textwidth]{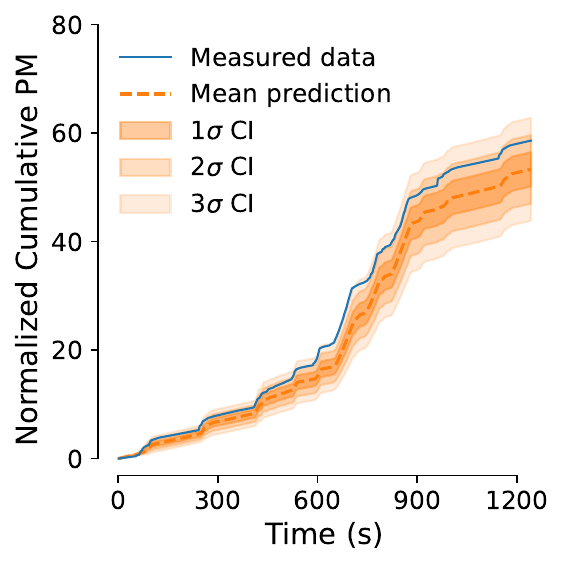}
        \caption{FTP - Low NOx}
        \label{fig:cum_ftp}
    \end{subfigure}
    \begin{subfigure}[b]{0.3175\textwidth}
        \centering
        \includegraphics[width=\textwidth]{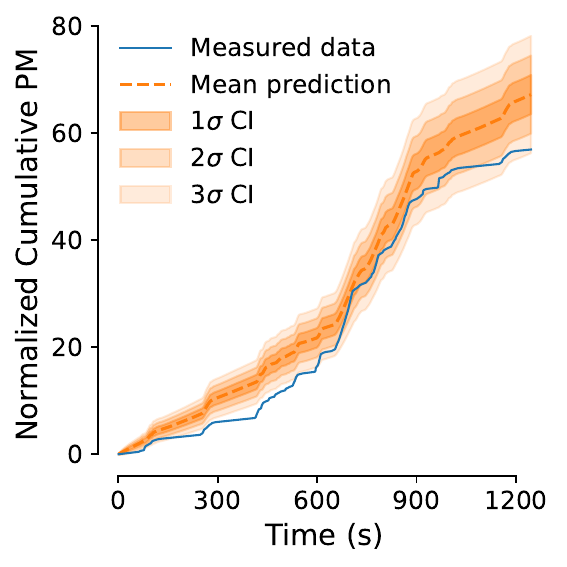}
        \caption{FTP - Nominal NOx}
    \end{subfigure}
    \begin{subfigure}[b]{0.3175\textwidth}
        \centering
        \includegraphics[width=\textwidth]{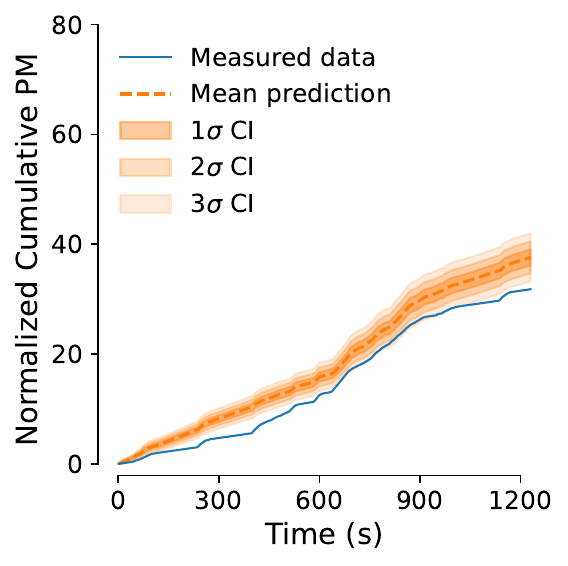}
        \caption{FTP - High NOx}
    \end{subfigure}
    \hfill
    \begin{subfigure}[b]{0.3175\textwidth}
        \centering
        \includegraphics[width=\textwidth]{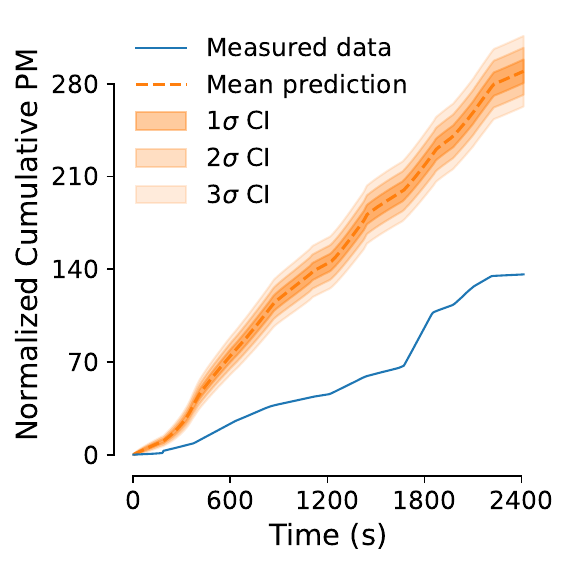}
        \caption{RMCSET - Nominal NOx}
        \label{fig:cum_rmcset}
    \end{subfigure}
    \begin{subfigure}[b]{0.3175\textwidth}
        \centering
        \includegraphics[width=\textwidth]{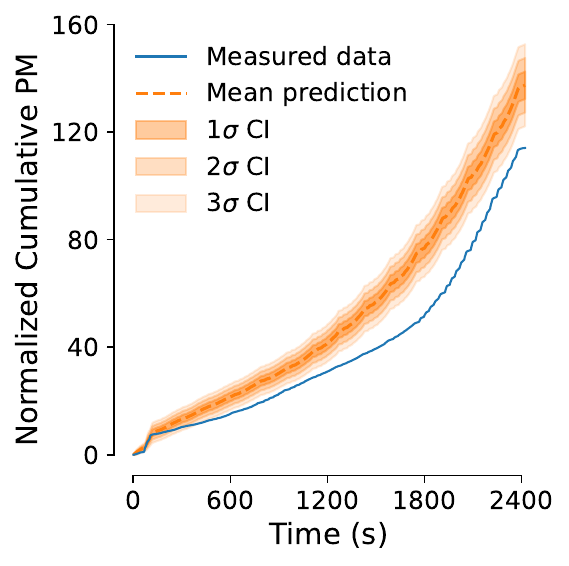}
        \caption{STC - Nominal NOx}
    \end{subfigure}
    \begin{subfigure}[b]{0.3175\textwidth}
        \centering
        \includegraphics[width=\textwidth]{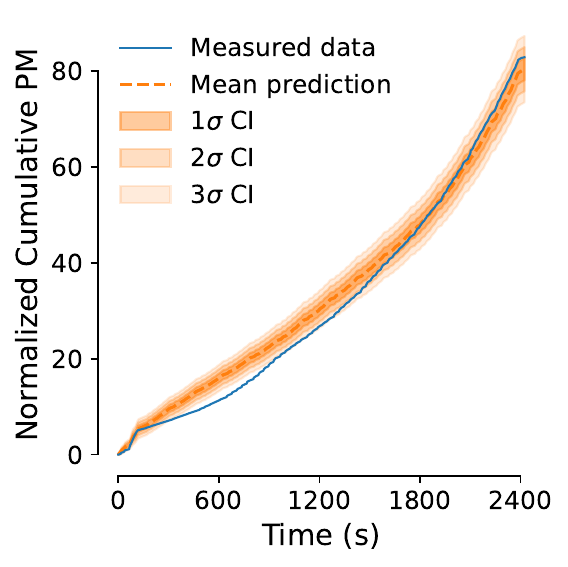}
        \caption{STC - High NOx}
    \end{subfigure}
    \caption{Cumulative PM output for test datasets.}
    \label{fig:cumulative}
\end{figure*}

Another metric of success that we have set for our models is whether or not the measured data stays within the CI throughout the duty cycle.
This can be checked by comparing the cumulative output of PM across time, where for each time step $t_k$, this is
$$
\text{Cumulative PM at $t_k$} = \sum_{i=1}^k X_{t_i},
$$
for any given sample path $X_t$.
We calculate the cumulative PM mass individually for each sample path and record the statistics.
Specifically, we report the mean and three CIs, representing one, two, and three standard deviations from the mean, for all six test datasets in Fig.~\ref{fig:cumulative}.
The model meets this criterion for three of the six test datasets.
Despite the wide CI of the RMCSET time series plot shown in Fig.~\ref{fig:RMCSET_ts}, the CI of the cumulative output of the RMCSET is narrower than the FTP cumulative PM.
This indicates that although each sample is very volatile, each sample has similar volatility, resulting in a narrow cumulative output.

\subsection{Appending the training data}
\label{sec:appending-training-data}
In an effort to improve the model performance on the RMCSET test dataset, we experimented with adding the first half of the RMCSET to the training and testing the model on the remainder. 
We found this not only improved testing performance on the RMCSET cylce but also improved model performance on every test dataset except for FTP - Low NOx.
We record the resulting KS statistics in Table~\ref{tab:appended-ks}.

\begin{table}[!htbp]
  \centering
  \caption{Kolmogorov–Smirnov statistics for RMCSET appended training.}
  \label{tab:appended-ks}
  \begin{tabular}{lcc}
    \toprule
    \textbf{Test Dataset} & \textbf{Original} & \textbf{Appended} \\
    \midrule
    FTP - Low NOx        & 0.085               & 0.139 \\
    FTP - Nominal NOx    & 0.290               & 0.196 \\
    FTP - High NOx       & 0.246               & 0.239 \\
    RMCSET - Nominal NOx & 0.482               & 0.390 \\
    STC - Nominal NOx & 0.134               & 0.064 \\
    STC - High NOx    & 0.057               & 0.045 \\
    \bottomrule
  \end{tabular}
\end{table}

\subsection{Benchmarking}
\label{sec:benchmarking}
We show our OU model benchmarked against six different model architectures of various kinds directly comparing two GPs, three neural networks, and one regression tree. 
Note that we do not include SINDy or a neural ODE model as their performance metrics are magnitudes worse than the selected architectures.
To ensure fairness between models, we performed hyperparameter optimization and early stopping with the same validation dataset used to tune the OU model. 
Below, we describe the model selection process for each broader category.

We use two different GPR models implemented using the GPyTorch library~\cite{GPyTorch}. 
Due to the size of the dataset, exact GPR is not feasible. 
Instead, we use approximate GPR both with a simple kernel, which we call the standard GP, and a deep kernel. Both GPs are equipped with a scaled RBF covariance kernel and trained with the Adam optimizer~\cite{kingma2017}.
For the standard GP, we searched over both constant and linear mean functions, trying combinations with 10, 20, 50, 100, 200, 500, 1000, and 1500 inducing points. 
The large range in number of inducing points resulted from the hyperparameter search continually showing us that smaller GPs with aggressive learning rates performed better in validation. 
We experimented in several stages with over 120 configurations to reach the best performing GP which has linear mean, 50 inducing points, a learning rate of 0.1, and a batch size of 5000.

Because of the success modeling NOx and the similarity in physical processes between NOx and PM, we implemented an approximate GP with the deep kernels described by Zinage et al.~\cite{zinage2025causal}. We explored using three different types of deep kernels, fully connected networks (FCNs), convolutions neural networks (CNNs), and graph convolutional networks (GCNs). The data passed through the deep kernel is projected to a learned latent space and then input to a zero mean GP. To reduce the hyperparameter search space, we used the CNN and GCN deep kernel architectures described by Zinage et al. Lacking a specified FCN architecture, we did some preliminary experimentation and arrived at four hidden layers with 256, 128, 64, 32 neurons each. Freezing these architectures, we trained over 500 deep kernel GPs experimenting with 500, 1000, 1500 inducing points, a window size of 5, 10, 20 time steps of engine state and control history, learning rates of 0.01, 0.001, and 0.0001, and batch sizes of 64, 256, and 512 data points. For the FCN and CNN we experimented with deep kernel output dimensions of 3, 12, and 20. The output dimension determines the dimension that the approximate GP sees and can be thought of as the number of features extracted from the window of input data. For the prescribed GCN architecture the output dimension was frozen at 12. Despite the success of the GCN deep kernel with NOx, we found that the CNN deep kernel performed better in validation than both the FCN and GCN deep kernel. After searching over 500 different configurations, the best performing model is a CNN deep kernel with an output dimension of 20, 1500 inducing points, trained with 5 time steps of history with a learning rate of 0.01 and a batch size of 64. The test performance of the deep kernel GP is universally worse than the standard GP, indicating that the deep kernel GP is likely overfitting beyond what can be mitigated with early stopping or that PM dynamics are more Markovian than NOx dynamics. 

We compare our OU model and the GPs with three different neural network forms all implemented in PyTorch~\cite{PyTorch} and trained with the Adam optimizer. 
First, we looked at simple neural networks with a hyperbolic tangent activation function and input the current engine variable time step. 
We explored total of 138 total hyperparameter configurations with 2-4 hidden layers whose widths stayed constant or decreased toward the output layer. 
Each of these configurations was explored with dropout rates of 0\%, 10\%, and 20\% and a learning rate of 0.001, 0.003, or 0.0001. 
The best performing simple neural network had three hidden layers of 256, 128, and 64 neurons respectively. 
It was trained with a learning rate of 0.003, dropout rate of 0\%, and a batch size of 512.

The second neural network form is a long short-term memory (LSTM) network~\cite{LSTM} designed to capture non-Markovian effects by including a window of past engine variables, but not the past PM predictions. 
During several stages of hyperparameter optimizations where we trained over 400 models, we noted that validation error continued to slowly shrink as network size increased. 
We capped our exploration at a network that observed 40 previous time steps with 5 hidden layers each having 1024 neurons. 
This network is already prohibitively large for embedded applications, while the validation error remained larger than other benchmark models. 
The recorded model was trained with a learning rate of 0.001 and a batch size of 512.

For the third neural network based architecture, we implemented a nonlinear autoregressive exogeneous (NARX) model~\cite{NARX}. 
The NARX model inputs the past PM prediction time steps and the current engine state to a neural network to predict the next PM time step. 
During hyperparameter optimization we explored 96 different models. 
They ranged from 2-4 hidden layers of progressively non-increasing neuron sizes. 
We explored a window of either 5, 10, or 20 past PM time steps to feed into the model. 
We trained each model with a learning rate of 0.01, 0.001, and 0.0001; and batch size of either 256 or 512 training points. 
Our best NARX model has three layers of 256, 256, 64 neurons, a window size of 10, a learning rate of 0.001, and a batch size of 512. 

Finally, we selected gradient boosted regression trees~\cite{GradBoost} as the  best regression tree algorithm when compared with random forest, support vector regression, and histogram-based gradient boosted regression trees. 
An extensive hyperparameter search was done on gradient boosted regression trees, searching a maximum depth of 2, 3, 4, or 5 with 400, 800, 1000, or 2000 estimators each. 
We tried learning rates of 0.3, 0.1, and 0.05 for each configuration. 
The best performing model had a maximum depth of 3 with 1000 estimators each and a learning rate of 0.1.
The regression tree models were implemented in scikit-learn~\cite{scikit-learn}. 

We use two different metrics to benchmark the models. 
First, we show the normalized root mean square error (NRMSE). 
The majority of the models that we use as benchmarks do not predict distributions, so NRMSE gives us a way to quantify point prediction errors.
We see in Table~\ref{tab:nrmse_benchmark} that the OU model outperforms all other models in three of the six datasets, a fourth it ties with the standard GP and LSTM, and the remaining two are split between the standard GP and the simple neural network.
\begin{table*}[!htbp]
  \centering
  \caption{NRSME for all benchmark models across six datasets.}
  \label{tab:nrmse_benchmark}
  \begin{tabular}{lccccccc}
    \toprule
    Dataset & OU & Standard GP & Deep kernel GP & NN & LSTM & NARX & Grad.\ boosted \\
    \midrule
    FTP - Low NOx         & \textbf{0.052} & 0.060 & 0.066 & 0.059 & 0.065 & 0.064 & 0.065 \\
    FTP - Nominal NOx     & \textbf{0.059} & 0.061 & 0.069 & 0.061 & 0.070 & 0.070 & 0.070 \\
    FTP - High NOx        & \textbf{0.015} & \textbf{0.015} & 0.016 & 0.019 & \textbf{0.015} & 0.017 & 0.018 \\
    RMCSET - Nominal NOx  & 0.072          & \textbf{0.049} & 0.057 & 0.059 & 0.055 & 0.053 & 0.055 \\
    STC - Nominal NOx  & \textbf{0.054} & 0.057 & 0.060 & 0.057 & 0.065 & 0.060 & 0.063 \\
    STC - High NOx     & 0.030          & 0.032 & 0.031 & \textbf{0.029} & 0.035 & 0.031 & 0.032 \\
    \bottomrule
  \end{tabular}
\end{table*}
Second, we benchmark the probabilistic models with the KS statistic of the PIT values as described above. 
We see in Table~\ref{tab:ks_benchmark} that the OU model scores the lowest KS value on four out of the six datasets with the standard GP scoring the lowest on the remaining two datasets.

\begin{table*}[!htbp]
  \centering
  \caption{KS statistics for probabilistic benchmark models.}
  \label{tab:ks_benchmark}
  \begin{tabular}{lccc}
    \toprule
    Dataset & OU model & Standard GP & Deep kernel GP \\
    \midrule
    FTP - Low NOx       & \textbf{0.085}    & 0.191             & 0.374 \\
    FTP - Nominal NOx   & \textbf{0.290}    & 0.354             & 0.488 \\
    FTP - High NOx      & 0.246             & \textbf{0.221}    & 0.356 \\
    RMCSET - Nominal NOx & 0.482            & \textbf{0.324}    & 0.522 \\
    STC - Nominal NOx & \textbf{0.134}   & 0.199             & 0.472 \\
    STC - High NOx   & \textbf{0.057}    & 0.259             & 0.452 \\
    \bottomrule
  \end{tabular}
\end{table*}

In Table~\ref{tab:avg_metrics} we present the average of each metric across each of the test datasets. 
We exclude the RMCSET data, treating it as an outlier due to our analysis that determined it is out of distribution. 
We see that our OU model on average outperforms all other models in both NRMSE and KS statistic. 
Additionally, we compare the total number of model parameters for each model, noting that the OU model is at least two orders of magnitude smaller than all of the benchmark models.

\begin{table}[htbp]
  \centering
  \caption{Benchmark models average NRMSE and KS statistics across five datasets (excluding RMCSET)}
  \label{tab:avg_metrics}
  \begin{tabular}{lccc}
    \toprule
    Model & \# Params & Avg.\ NRMSE & Avg.\ KS \\
    \midrule
    OU model        & \textbf{35}   & \textbf{0.042} & \textbf{0.162} \\
    Standard GP     & 3385          & 0.045          & 0.248          \\
    Deep kernel GP  & 2433185       & 0.048          & 0.428          \\
    NN              & 46465         & 0.045          & --             \\
    LSTM            & 37856257      & 0.050          & --             \\
    NARX            & 89217         & 0.048          & --             \\
    Grad.\ boosted  & 21446         & 0.049          & --             \\
    \bottomrule
  \end{tabular}
\end{table}

\section{Conclusions}
\label{sec:conclusions}
In this work, we introduced SDE-based models for modeling dynamics with high-dimensional inputs. 
By using the OU process form, we model the delay from engine state measurement to the downstream PM measurement.
We parametrize the SDE terms and fit the parameters with the maximum likelihood principle. Using a synthetic example, we verify the model's ability to capture time-varying dynamics. 
We then implement our model for PM with linear parameterizations of both the mean and volatility functions, with the engine state as input.

Our OU model on average outperforms every regression method that we applied to the PM problem and addresses the systematic errors that we observed in our preliminary regression models. 
Our framework enables us to generate a distribution rather than a point prediction which is essential to capture the stochastic behavior that we observe in PM. 
The model's worst performance is on predicting the transition between steady-state operating conditions in the RMCSET data. 
We showed that the performance significantly improved when we introduced half of the RMCSET test data as training data. 
Despite the efforts to capture all possible operating conditions by dynamically moving through the engine control space, we conclude that our PM model would benefit from the collection of new data that transitions from one steady-state operating point to another steady-state operating point.
 
To further improve the model, we will explore increasing the model complexity including the parametrization of the drift and volatility terms. 
We explored neural network parametrization, but these led to excessive overfitting beyond what could be mitigated with early stopping. 
Another way we attempted to introduce model nonlinearity was mapping PM with a neural ODE~\cite{NODE} to a space where it follows the prescribed OU process.
But this transformation also led to the same overfitting problems that we experienced with other neural network parameterizations. 
A fully Bayesian treatment would enable uncertainty propagation for $\mu_t$, $\sigma_t$, and $\lambda$ and support informative priors reflecting known constraints (e.g., positivity and temporal smoothness). 
Hierarchical Bayesian formulations provide a principled way to encode such structure and quantify uncertainty under limited observations~\cite{hans2023bayesian}, while scalable variational inference offers a practical route for Bayesian training in high-dimensional likelihood settings~\cite{hans2024bayesian}.

Although there are many ways to increase model complexity, this current model consists of only 35 parameters and is well suited for deployment on limited-memory engine control modules. Our OU model provides improved engine-out PM prediction for emissions compliance and diagnostics in the next generation of diesel engines.

\section*{Author Contributions}
\noindent\textbf{Maxwell Bolt:} Software, Methodology, Validation, Visualization, Writing -- original draft.\\
\textbf{Alex Alberts:} Validation, Writing -- original draft.\\
\textbf{Akash S. Desai:} Data curation, Writing -- review and editing.\\
\textbf{Peter Meckl:} Funding acquisition, Validation, Writing -- review and editing.\\
\textbf{Ilias Bilionis:} Funding acquisition, Methodology, Validation, Writing -- review and editing.

\section*{Acknowledgments}
The authors would like to thank Lisa Farrell and Clay Arnett at Cummins Inc.\ for their feedback and support.

\section*{Conflict of Interest}
The authors declare that there are no conflicts of interest related to the research, authorship, or publication of this article.

\section*{Funding}
This work has been funded by Cummins Inc.\ under proposal number 00099056.

\bibliographystyle{unsrt}
\bibliography{references}

\end{document}